# Runtime Safety Assurance of Autonomous Vehicles used for Last-mile Delivery in Urban Environments

*Iqra Aslam, Adina Aniculaesei, Abhishek Buragohain, Daniel Bamal, Prof. Dr. Andreas Rausch*

*Institute for Software and Systems Engineering, TU Clausthal*

*{iqra.aslam, adina.aniculaesei, abhishek.buragohain, daniel.bamal, andreas.rausch}@tu-clausthal.de*

**Abstract**

Last-mile delivery of goods has gained a lot of attraction during the COVID-19 pandemic. However, current package delivery processes often lead to parking in the second lane, which in turn has negative effects on the urban environment in which the deliveries take place, i.e., traffic congestion and safety issues for other road users. To tackle these challenges, an effective autonomous delivery system is required that guarantees efficient, flexible and safe delivery of goods. The project LogiSmile, co-funded by EIT Urban Mobility, pilots an autonomous delivery vehicle dubbed the Autonomous Hub Vehicle (AHV) that works in cooperation with a small autonomous robot called the Autonomous Delivery Device (ADD). With the two cooperating robots, the project LogiSmile aims to find a possible solution to the challenges of urban goods distribution in congested areas and to demonstrate the future of urban mobility. As a member of Niedersächsische Forschungszentrum für Fahrzeugtechnik (NFF), the Institute for Software and Systems Engineering (ISSE) developed an integrated software safety architecture for runtime monitoring of the AHV, with (1) a dependability cage (DC) used for the on-board monitoring of the AHV, and (2) a remote command control center (CCC) which enables the remote off-board supervision of a fleet of AHVs. The DC supervises the vehicle continuously and in case of any safety violation, it switches the nominal driving mode to degraded driving mode or fail-safe mode. Additionally, the CCC also manages the communication of the AHV with the ADD and provides fail-operational solutions for the AHV when it cannot handle complex situations autonomously. The runtime monitoring concept developed for the AHV has been demonstrated in 2022 in Hamburg. We report on the obtained results and on the lessons learned.

**Table of contents**

# 1      Introduction

The volume of e-commerce transactions has increased significantly in the past years, even more so during the COVID-19 pandemic, e.g., 4.05 billion shipments were carried out by the courier, express and parcel (CEP) services sector. More than half of the world's population currently lives in cities or urban agglomerations, and it is estimated that three quarters of the world's population will live in cities by 2050. Thus, many of the customers of CEP services are located in cities or in large urban areas. However, current solutions for the urban logistics sector have negative effects on the urban environment in which the deliveries take place, i.e., traffic congestion due to parking in the second lane, greenhouse emissions, and safety issues for other road users. In order to address these issues and make their urban centres more liveable for their citizens, many cities are looking at new concepts for the administration of a city. For example, Paris is implementing the model of the *15-minute city* (Gongadze and Maassen, 2023). In this model, every citizen lives within 15 minutes, including different means of transportation, from all the main urban facilities, e.g., supermarkets, restaurant, or sport centres as well as city services that he or she needs, e.g., public administration offices or hospitals. While these models make the design of the public space more user-centric, they also represent an additional challenge for the parcel logistics sector in large urban areas. City administrations aim to limit the number of vehicles in the city by increasing the number of pedestrian areas, limiting the number of parking spots and charging congestion taxes to the vehicles.

The shortage of transport space as well as the increased regulation make it more difficult for operators in the CEP sector to carry out their deliveries in large urban centres. One consequence of this is that the delivery person spends little time with the actual delivery and the majority of his or her working time with driving the vehicle. Furthermore, the search of parking space is a time-consuming and therefore cost-intensive activity, which does not add value to the delivery process. Automation of the delivery process can be used to address this increased operative complexity in the urban logistics sector and make it more sustainable. The LogiSmile project[1] proposes an automated delivery system consisting of an autonomous delivery vehicle dubbed as the autonomous hub vehicle (AHV) and a small autonomous robot called as the autonomous delivery device (ADD), which cooperate with each other in order to carry out deliveries to customers in urban centres. The AHV is a full-size electrical car which is able to cover large distances on open streets in urban areas and can transport a large number of parcels. The ADD is a smaller autonomous robot which covers pedestrian areas in order to reach its end customers. Due to the large number of parcels it can transport, the AHV acts as a mobile depot for the ADD. In turn, the ADD picks up the parcels from the AHV and delivers them on the last mile to the end customer.

The AHV is planned as a vehicle of at least SAE Level 3, in which the automated driving system (ADS) is expected to take over the responsibility of the dynamic driving task partially or completely. In order to ensure the safety of the AHV, we propose an integrated safety concept for ADSs, which consists of a combination of onboard runtime monitoring via a connected dependability cage (DC) and off-board runtime monitoring through a remote command control centre (CCC). The DC monitors continuously the operation of the automated driving system and in case of a safety violation, it switches its nominal driving mode (full autonomous driving) to a degraded driving mode (limited autonomous driving) or to a fail-safe mode (emergency brake). In turn, the CCC enables a remote operator the monitoring and takeover of the vehicle's control, should a complex situation arise that the AHV cannot handle autonomously. Furthermore, the CCC enables the supervision of a fleet of AHVs and manages the communication of the AHV with the ADD, in order to synchronize the two autonomous vehicles for the last mile delivery. The integrated safety concept developed for the AHV is evaluated in an urban-like test environment in Hamburg using a use case scenario from the domain of parcel delivery logistics, which is defined together with academic and industry partners in the project LogiSmile.

---

[1] https://www.eiturbanmobility.eu/projects/logismile/

The rest of this paper is structured as follows. Section 2 gives an overview of current monitoring approaches for automated driving systems, while Section 3 presents our integrated safety concept for the runtime assurance of ADSs via the connected dependability cage approach. Section 4 discusses the results obtained during the evaluation in the urban-like test environment. Section 5 concludes this paper and points towards future research directions.

## 2   Related Work

In their work, Li et al. (2020) propose monitoring systems that aim to detect potential hazards and implement proactive measures to prevent accidents in autonomous driving, particularly in challenging traffic conditions (cf. (Li et al., 2020)). The study emphasizes the utilization of sensor fusion, computer vision, and real-time data analysis to monitor the vehicle's surroundings and make informed decisions for safe navigation. However, one limitation of their approach is the absence of human involvement during the driving process, making it difficult to ensure fail-safe operations. Kwon et al. (2022) underscore the critical role of monitoring approaches in web-based car control and monitoring for autonomous vehicles (cf. (Kwon et al., 2022)). Real-time monitoring systems, leveraging sensors and data processing techniques, continuously gather information to identify potential risks and enable prompt interventions. However, a challenge that arises is the reliance on human experts for remote monitoring, as they may not consistently pay attention to every single detail.

Ge et al. (2022) adopt a systematic approach using the STPA method for monitoring autopilot sensors in autonomous delivery vehicles (cf. (Ge et al., 2022)). Their real-time monitoring focuses on detecting anomalies and ensuring safe vehicle operations. However, their work primarily focuses on the sensor level, and there is a lack of comprehensive fail-safe mechanisms in their approach. Sherry et al. (2020) discuss monitoring approaches in the deployment of an autonomous shuttle bus, emphasizing the importance of real-time sensor data collection and fault detection for ensuring safe and reliable operations (cf. (Sherry et al., 2020)). Their research highlights the significance of robust monitoring strategies in the development and deployment of autonomous systems, contributing to the overall safety and reliability of autonomous transportation.

To address the limitations observed in the previous works, this paper builds upon the foundation of research presented in several prior publications. The concept of a dependability cage was initially introduced in Aniculaesei et al. (2018), which outlined the challenges related to engineering hybrid AI-based Advanced Driver Assistance Systems (ADASs) concerning dependability and safety assurance (cf. (Aniculaesei et al., 2018)). The application of the dependability cage concept was demonstrated in a lane change assistance system (LCAS) in (Mauritz, 2020). More recently, the concept of a connected dependability cage was introduced as an extension of the initial notion of a dependability cage in (Helsch et al., 2022). Its application in the context of parcel delivery logistics within the VanAssist project was detailed in a previous publication by the authors (cf. (Helsch et al., 2022)). Compared to a previous version of the safe zone computation presented in (Grieser et al., 2020), the focus in (Helsch et al., 2022) was on enhancing the algorithm for computing the safe zone around the ego-vehicle, addressing specific challenges encountered in the VanAssist project, e.g., handling ghost points in the LiDAR sensor data through specific clustering algorithms and differentiating between the relevant LiDAR data points and the background in a 3D space using a z-cutoff algorithm (cf. (Helsch et al., 2022)).

## 3    Runtime Assurance via Connected Dependability Cage Approach

Ensuring the safety of autonomous driving systems is a crucial both during development and operation time. In lower level of automation (SAE Levels 0 - 2), an onboard safety driver is always required to take control from the autonomous driving system in case a problem occurs. For higher levels of automation (SAE Levels 3 - 5), the safety driver is theoretically not needed anymore and therefore does not have to be onboard the vehicle anymore. However, even for higher levels of automation, legal regulations require that a safety driver be part of the process as a last fall-back mechanism, which is activated in case a complex situation cannot be handled autonomously by the vehicle. One way to ensure the benefits of autonomous driving is by using off-board safety drivers that can remotely control the whole fleet for vehicles by combining an onboard runtime monitoring with this remote control and human intervention (cf. (Aniculaesei et al., 2019)). To ensure the safety of ADSs, we propose an integrated safety concept consisting of two cooperative redundant systems: (1) an onboard runtime monitoring system, called DC, and (2) an off-board runtime monitoring system realized with a CCC and human remote operators. This approach is a generalized concept that addresses safety concerns in various automated mobility solutions. It has been specifically designed to meet the requirements of the research project LogiSmile, considering its specific use case. In the following we describe the integrated cooperative approach for safety in ADSs at runtime (cf. Section 3.1) as well as overall software architecture for ensuring safety, the internal structure of the DC and an overview of the CCC (cf. Section 3.2).

### 3.1    Integrated Cooperative Approach concept for Ensuring Safety in Autonomous Driving Systems at Runtime

Figure 1 provides an overview of our integrated cooperative safety concept to ensure the safety of ADS at runtime. This integrated safety concept, designed to ensure safety for ADSs, employs a redundant monitoring system that operates at different levels of abstraction within the autonomous vehicle. On one hand, the DC is specifically focused on guaranteeing the safety of a single autonomous vehicle during its runtime. To accomplish this, the DC analyses the raw sensors data received from the vehicle at runtime. It compares the sensor's data against the modelled safety specifications to determine if any specification violations exist (as depicted in the (a) upper section of Figure 1). On the other hand, the remote operator stationed in the CCC monitors an entire fleet of vehicles, extending beyond the scope of a single autonomous vehicle (as depicted in the (a) lower section of Figure 1).

When the DC detects any violation of the safety specification, it switches from the nominal driving mode to a fail-safe mode, e.g., through a request for an emergency stop, so that the autonomous vehicle can be brought to a safe state (as depicted in (b) upper section of Figure 1). Subsequently, the information about the identified problem is transmitted via wireless communication to the CCC. The remote operator in the CCC is able to visualize the live stream of the vehicle sensors data. Based on it, the remote operator assesses the current situation to confirm the problem case. Once confirmed, the remote operator takes a decision with respect to the corresponding human intervention (as depicted in (b) lower section of Figure 1). Through its human-machine interface (HMI), the CCC provides different options for human intervention, e.g., reconfiguration of the ADS or directly controlling the autonomous vehicle remotely. Notice that, in the second option, the remote operator can be regarded as an off-board safety driver.

If the identified problem cannot be resolved remotely by the remote operator in the CCC, then the remote operator requests local assistance, e.g., by a delivery person present nearby the autonomous vehicle. After the problem has been resolved, the remote operator releases autonomous driving for the vehicle again (as shown in (c) upper part of Figure 1), while continuing to monitor the entire fleet (as shown in (c) lower part of Figure 1).

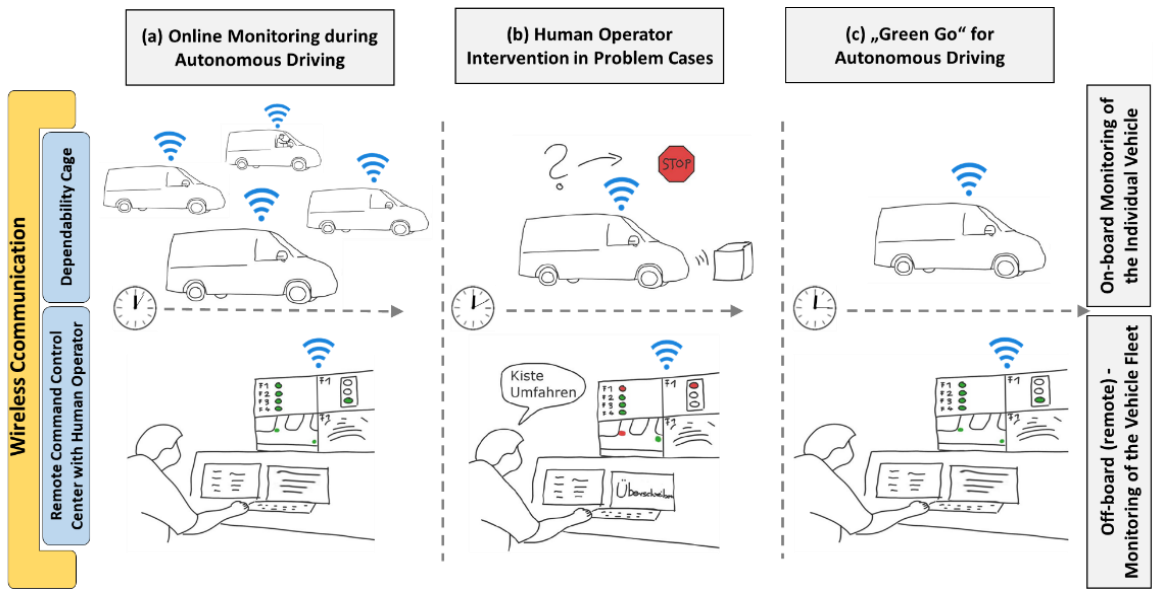

Figure 1: Overview of Integrated Cooperative Safety Concept for ADSs at Runtime (cf. (Seber et al., 2021)).

## 3.2   Integrated Cooperative Software Architecture for Ensuring Safety in Autonomous Driving Systems at runtime

Based on the above defined safety concept, the concrete functions of the DC and the CCC are formulated and an integrated cooperative software safety architecture for the entire cooperative safety solution is defined. Figure 2 illustrates the integrated software safety architecture for ADSs, which consists of three layers: (a) a reconfigurable modular ADS, as previously described by (Raulf et al., 2021b), which performs the driving task and incorporates reconfiguration interfaces in a computation pipeline that uses data from various sensors. (b) a DC, as explained by (Helsch et al., 2022), which is responsible for executing the onboard qualitative monitoring and fail-operational reaction at runtime for the individual vehicle and (c) a CCC with HMI interface, which allows for offboard supervision and human intervention by a remote operator. Notice that a system that allows remote operators to monitor and control autonomous vehicles is also required by German regulations. The German government has already drafted a bill to amend the German Road Traffic Act and the German Compulsory Insurance Act, and thus sets the legal framework for automated and autonomous driving systems (cf. (German Government, 2021)).

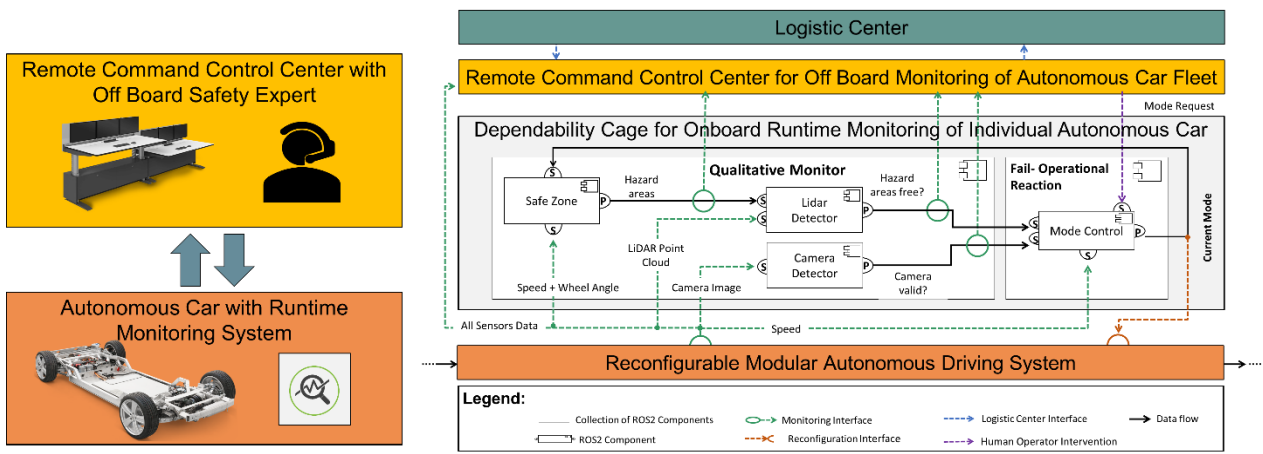

Figure 2: Integrated Safety Architecture for ADSs. (Everding et al., 2023).

The integrated safety concept was developed based on the collaboration between the DC and the CCC to ensure the safety of autonomous vehicle, as mentioned in the previous section. In this case, the DC operates autonomously without human intervention during its operation. On the other hand, the command control centre is an interactive system that involves a human remote operator. This collaboration between the two systems can be seen as a human-machine collaboration. The integrated safety concept supports the sharing of responsibilities between the autonomous vehicle and the human remote operator, thus emphasizing the role of the remote safety driver in maintaining the vehicle safety.

#### 3.2.1. Internal Structure of Dependability cage

Figure 2 illustrates in the middle right part the internal software component view of the DC. Originally, the DC has two main components: (1) the *qualitative monitor* (cf. (Aniculaesei et al., 2018)), that verifies system correctness based on safety specifications and (2) the *quantitative monitor* that detects unknown driving situations as novelties (cf. (Aniculaesei et al., 2018)). Since it was not in the project's focus, the quantitative monitor was not implemented as the part of project LogiSmile.

For the **qualitative monitor**, some concrete safety specifications were identified and implemented. It consists of three components: Safe Zone, Lidar Detector, and Camera Validator. The *Safe Zone* calculates the hazard-free area around the vehicle considering the vehicle dynamics, the current velocity, and the steering angle. The *Lidar Detector* utilizes LiDAR data points from the vehicle's LiDAR sensors and the results of the safe zone computation to check for obstacles within the safe zone. Lastly, the *Camera Validator* component verifies the validity of images received by the ADS from the camera sensors. Further technical details regarding each of these components can be found in (Helsch et al., 2022), which gives an in-depth analysis of the respective functionalities and specifications of each components.

The **fail-operational reaction** is determined by a *Mode Control* component, which take as inputs the results of the Lidar Detector and the Camera Validator, the speed and the steering angle as well as the requests for change of the current cage mode and of the driving mode received from the CCC and computes the current driving mode. Depending on the inputs received by the Mode Control component, the computed driving mode could be either the nominal driving mode, the degraded driving mode, or the fail-safe mode. Once a fail-safe mode occurs, only remote operator (via the HMI of the CCC) has the authority to select the next appropriate driving mode, e.g., a degraded driving mode such as limited autonomous driving. The Mode Control component then triggers the selected mode based on the current inputs received from the qualitative monitor. This process of responsibility transfer between the human operator and the autonomous vehicle is implemented through the concept of graceful degradation (cf. (Aniculaesei et al., 2019)). This ensures a smooth transition and adaptation of the system in response to failures. Based on the extensive literature search by (Warnecke et al., 2018), each component in this software architecture is developed using the middleware ROS2[2], which has a completely decentralized architecture. The usage of ROS2 is further motivated by the fact that automotive systems are typically designed as distributed systems deployed on various ECUs.

#### 3.2.2. Remote Command Control Centre

The CCC allows the human remote operator to visualize the state of the autonomous vehicle based on the sensor data received from the LiDAR and camera sensors of the vehicle as well as the inputs received from the dependability cage. Figure 3 shows an overview of the graphical user interface (GUI) of the CCC. The widget for car selection on the left side of the command control centre displays a summary that includes: sensor data, mission state, door state, driving mode, and cage state.

The *sensor data* specifically refers to the validity of the data obtained from the front camera. The *mission state* distinguishes between several states: states inactive, active, blocked and completed. The *inactive* state means that the vehicle is not currently performing a driving task. The state *active* means that the vehicle is currently processing a driving task, which is has not yet been completed. If a problem occurs during the processing of a driving task ("Fail-safe Mode"), which prevents the vehicle from completing the current driving task, the state *blocked* is inferred. After the vehicle has completed its driving task, the *mission state* is considered to be completed. The *door state* reflects the status of the electrical rolling gate of the test vehicle PLUTO, while the *cage state* describes the current condition of the dependability cage. Collectively, these attributes provide an overall representation of the state of the test vehicle PLUTO. The possible values of each attribute are listed in Table 1.

---

[2] https://docs.ros.org/en/foxy/index.html

Table 1: Information with respect to the Current Vehicle State of PLUTO depicted in the CCC.

| Attribute Name | Attribute Values |
|---|---|
| **sensor data** | {valid, invalid} |
| **mission state** | {inactive, active, blocked, completed} |
| **door state** | {open, closed, no data} |
| **driving mode** | {fully autonomous driving, limited autonomous driving, remote manual driving, in-place manual driving} |
| **cage state** | {safe zone free. safe zone occupied} |

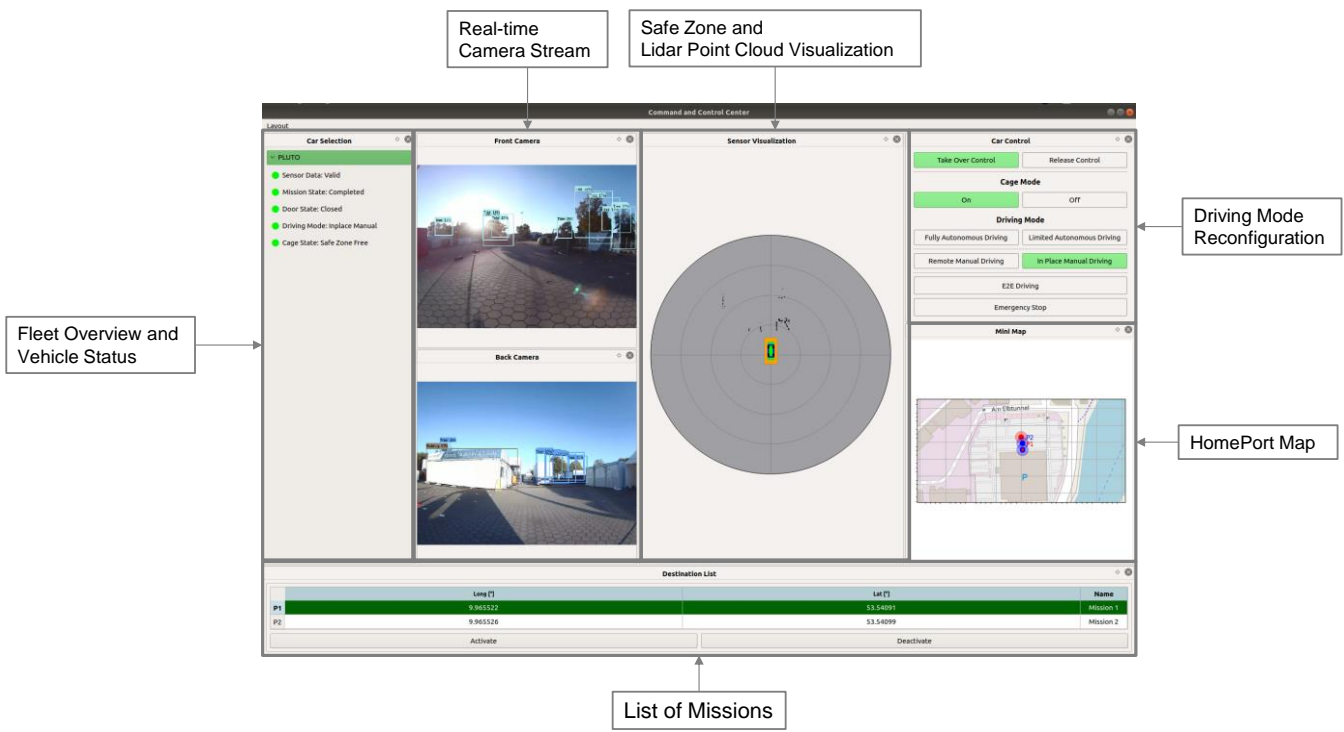

Figure 3: Overview of the GUI of the CCC.

Next to the car selection widget, there are two camera widgets named "front camera" and "back camera" that display the visual data captured by the respective cameras on the test vehicle PLUTO. The front camera widget shows the front view, while the back-camera widget displays the rear view. The LiDAR sensor data, along with the computed safe zone around the vehicle, is visualized in the sensor visualization widget. This provides a representation of the surrounding environment and helps ensure safe operation of the test vehicle. Located in the top right corner, the car control widget presents various controls, including car control, cage mode, and driving mode. These controls allow for the manipulation of the vehicle, the configuration of the DC, and the selection of the desired driving mode.

In the centre right corner of the display, a mini-map of the test field at homePORT in Hamburg is presented. This provides an overview of the test area and aids in navigation and orientation. At the bottom of the CCC display, a list of the missions is shown. This list provides information on the tasks that the autonomous vehicle is assigned to complete, e.g., driving to the specified destination.

## 4    Evaluation and Discussion of Results

In this section we present and discuss of the evaluation carried out at homePORT in Hamburg. Section 4.1 gives an overview of the AHV and describes the hardware and communication setup between the AHV and the CCC. Section 4.2 outlines the demonstration track. This is followed by a presentation of the scenario selected for showcasing the developed safety concept and a discussion of the results in Section 4.3.

### 4.1    Description of the AHV and of the Hardware and Communication Setup

Figure 4 illustrate the test vehicle AHV used in the demonstration, called PLUTO, which was developed by NFF. PLUTO is based on an electrically driven, fully automated chassis, denoted as motion board[3], which has been used since 2020 in different research project by various research groups to showcase their results.

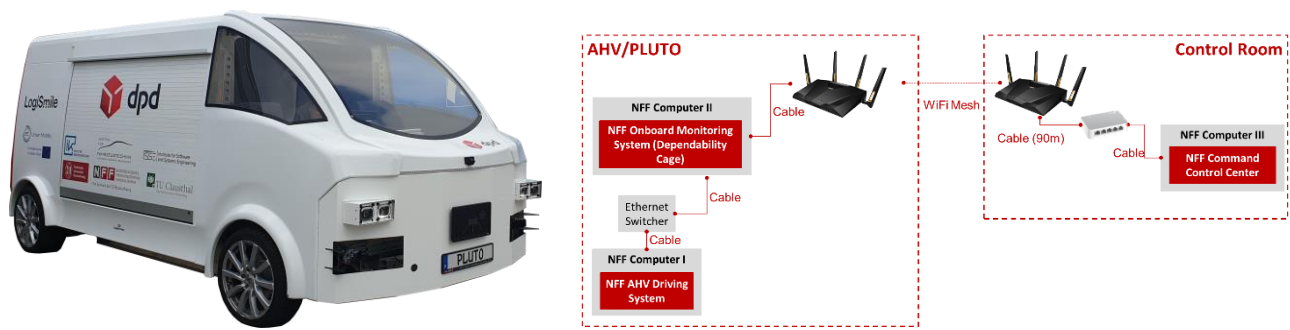

Figure 4: Test Vehicle (left) - Pluto and Communication Setup (right).

PLUTO is equipped with two Solid State Ibeo NEXT 3D LiDAR and two wide-angle fish-eye cameras. The LiDAR sensors are positioned at the front, while the cameras are mounted at the front and rear (see Figure 4). Additionally, a GeneSys ADMA Slim sensor is used for the purpose of localization. PLUTO's driving system was deployed on NFF PC I. An Industrial PC (NFF computer II) is mounted inside the PLUTO (cf. Figure 4, right), on which the DC is deployed for the onboard monitoring of the single vehicle using information from the mounted sensors. Furthermore, a second PC (NFF computer III) is placed in the control room, and is used to deploy the CCC in order to enable real-time off-board status monitoring for individual vehicles and for a fleet of vehicles as a whole. Figure 4 also illustrates the communication infrastructure put in place for the validation of developed safety concept. With the exception of the CCC, all other software components including DC runs on the Industrial PC NFF Computer II (see Figure 4 right). This industrial PC is connected to the main AHV Computer (NFF computer I) on which the ADS is deployed.

The NFF computer III, used to deploy the CCC, is connected with an ASUS RT-AX88U router which forms its own local Wi-Fi network. Another ASUS RT-AX88U router is also placed in the AHV which remains connected to the NFF computer II through a wireless network adapter on it. The two routers are connected to each other through a Wi-Fi Mesh Network so that there is a seamless connection for data exchange between the two systems of the integrated cooperative safety concept, the DC and the CCC**.**

---

[3] https://www.tu-braunschweig.de/nff/notiz-blog-detailansicht/autonom-dank-pluto

## 4.2　Description of the Test track

The test track is located at Hamburg, Germany belonging to homePORT[4]. An aerial view of the track is shown in Figure 5. The test track is part of a parking area site, which is enclosed from all sides (cf. Figure 5, right). The surrounding area of homePORT is used as parking space, so it was safe from the perspective of demonstrating on an automated driving vehicle. The site also had a coverage of 5G communication as well as good GPS reception, which were also one of the many requirements for our demonstration purposes, since the network communication for data exchange between PLUTO's NFF computers and the CCC Computer is over a WIFI mesh. Additionally, it also had a separate room which was used as the control room that housed the CCC and the human supervisor (cf. Figure 5, left). A spectator area situated above the control room allowed the public to view live the entire demonstration.

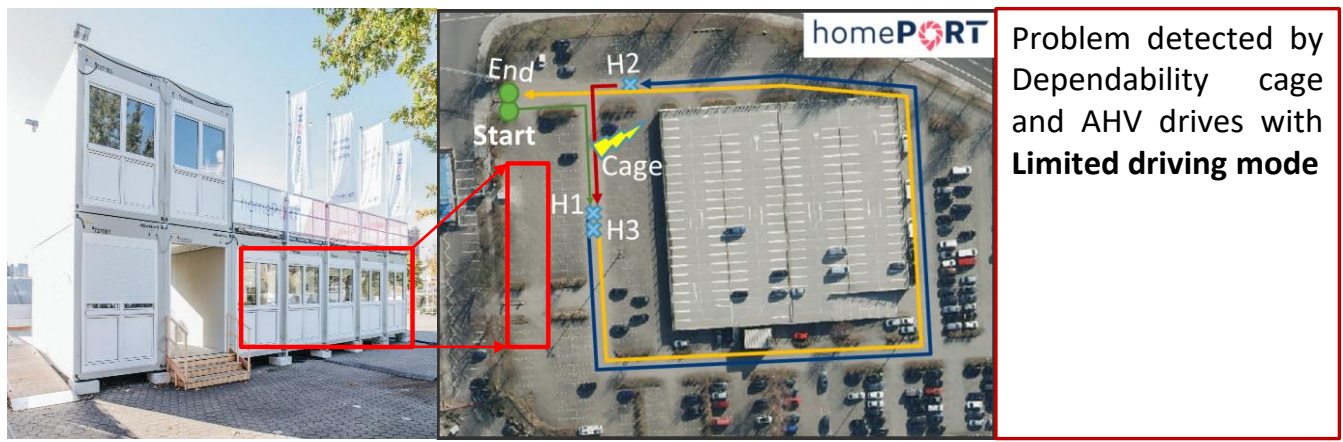

Figure 5:　Location of the Control Room with the CCC (left). Arial View of the Test Track (right).

## 4.3　Use Case Scenario and Results

The use case scenario selected for the test track pertains to the application domain of last-mile delivery of goods, in which different parcels must be delivered from the depot to various end customers situated in different delivery areas. The scenario was carried out according to the following steps:

(1) The postman loads the PLUTO with the parcel and PLUTO drives autonomously from *Start* position and reaches the delivery points *H1*, *H2* and *H3* in this order (cf. Figure 5, right).
(2) Once PLUTO has reached a delivery point, the door of PLUTO opens and the postman takes the corresponding parcel out of the car.
(3) Then the door of PLUTO is closed and the postman goes to deliver the parcel to the end customer.

In order to demonstrate our integrated safety concept consisting of the collaborative systems DC and CCC, we placed an obstacle on the road between H2 and H3, which narrows the path of the test vehicle. The obstacle is a vehicle driven by a human driver that is partially parked very close to the planned route of the PLUTO as shown in left part of the Figure 6, making it impossible for the test vehicle to drive in *Fully Autonomous Driving (FAD)* mode.

When the test vehicle drives on the track, the DC detects a parked vehicle as potential threat for the test vehicle. Since this is a new situation that has potentially not been considered during the vehicle development, the Mode Control component triggers the fail-safe mode *Emergency Stop (ES)* (cf. Figure 5, middle). The transition from the FAD mode to the ES mode causes the vehicle to come to a stop at a safe distance of $1\ m$ from the parked vehicle.

---

[4] https://www.homeport.hamburg/en/spaces-2

This problem is reported to human operator in the control room via wireless communication between the industrial PC of the DC and the CCC computer. The human operator assesses the situation and resolves it by changing the current driving system from ES to *Limited Autonomous Driving (LAD)* (cf. Figure 5, right). The difference between FAD and LAD modes is that LAD considers a smaller safety zone around the vehicle with limited speed compared to FAD. After switching to LAD, the test vehicle is able to successfully drive through the narrow path towards its next destination H3 for the package delivery. This demonstration showcases the developed integrated safety concept through graceful degradation and human intervention. The results of evaluation are also available as video[5], accessible to the larger public.

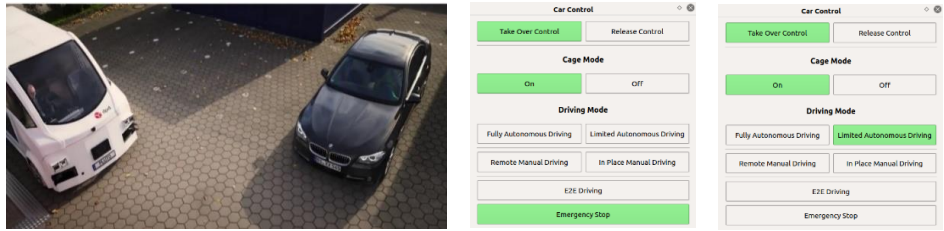

Figure 6: Demonstrated scenario (left). Reconfirmation of driving system ES (middle) and LAD (right) on the CCC.

## 5    Conclusion and Future Work

This paper presented an integrated safety concept for safeguarding the safety of ADSs based on the connected dependability cage approach. This approach consists of two runtime monitoring systems, which are integrated with each other and work in a collaborative manner: (1) a DC which monitors the ADS onboard the ego-vehicle and (2) the CCC which is able to supervise off-board an entire fleet of AVs with the cooperation of a remote safety driver. At the moment, when the DC detects an obstacle in the path of the ego-vehicle, it triggers immediately an emergency brake. This is a fail-safe reaction of the ego-vehicle. In future work, we plan to improve the DC so that it allows also fail-operational reactions. A possible fail-operational reaction could be similar to the reaction of a human driver, who in this case, would steer the vehicle backwards, go around the obstacle and then continue to drive. With respect to the domain of last-mile logistics, further improvements can be made in future research. In the demonstration carried out in Hamburg, the test vehicle interacted with a postman that took over the parcels from the vehicle in order to deliver them to the end customer. In future, we envision a collaboration between the AHV and another autonomous robot which plays the role of the postman and covers the last mile of the delivery process up to the door of the end customer. The testing and evaluation of our integrated safety concept took place in the closed and controlled environment of homePORT in Hamburg. In future, we plan to test this concept with the test vehicle in a use case with real driving conditions on a public road. This means that the safety concept must be adapted so that it can master the extensive interaction of the test vehicle with vulnerable road users and other vehicles as well as the various traffic rules that are enforced on the public road.

**Acknowledgements:** The authors would like to acknowledge the participation of all LogiSmile partners in the project and the EIT Urban Mobility for its support. The LogiSmile project was co-funded by the EIT UM under the KAVA 22140. Additional funding was provided from state funds available in the state budget of TU Clausthal provided by the state of Lower Saxony, Germany.

---

[5] https://www.youtube.com/watch?v=Y205SrL-jtM&t=59s

# 6     References

Aniculaesei, A.; Grieser, J.; Rausch, A.; Rehfeldt, K.; Warnecke, T. (2018): "Towards a holistic software system engineering approach for dependable autonomous systems," in Proceedings of the 1st International Workshop on Software Engineering for AI in Autonomous Systems, R. Stolle, S. Scholz, and M. Broy, Eds. New York, NY, USA: ACM, 2018, pp. 23–30.

Aniculaesei, A.; Grieser, J.; Rausch, A.; Rehfeldt, K.; Warnecke, T. (2019): "Graceful degradation of decision and control responsibility for autonomous systems based on dependability cages," in 5th International Symposium on Future Active Safety Technology toward Zero, Blacksburg, Virginia, USA.

German Government (2021): Draft law amending the road traffic act and the compulsory insurance act - autonomous driving act. Federal Ministry for Digital and Transport. Online verfügbar unter: https://www.bmvi.de/SharedDocs/DE/Anlage/Gesetze/Gesetze-19/gesetz-aenderung-strassenverkehrsgesetz-pflichtversicherungsgesetz-autonomes-fahren.pdf?__blob=publicationFile

Raulf, C.; Sahin, T.; Yarom, O. A.; Hegerhorst, T.; Pethe, C.; Aslam, I.; Zhang, M.; Vietor, T.; Lui-Henke, X.; Henze, R.; Rausch, A. (2021b): "Dynamically configurable vehicle concepts for passenger transport," In: 13th Wissenschaftsforum Mobilität:" Transforming Mobility – What Next", Duisburg, Germany. Accepted on May 2021.

Everding, L., Aslam, I., Raulf, C., Aviv Yarom, O., Fritz, J., Jacobitz, S., ... & Henze, R. (2023). Dynamically Configurable Autonomous Vehicles for Urban Cargo Transportation. In Towards the New Normal in Mobility: Technische und betriebswirtschaftliche Aspekte (pp. 851-869). Wiesbaden: Springer Fachmedien Wiesbaden.

Helsch, Felix; Aslam, Iqra; Buragohain, Abhishek; Rausch, Andreas (2022). "Qualitative monitors based on the connected dependability cage approach;" in Proceedings of the 14th International Conference on Adaptive and Self-Adaptive Systems and Applications - Barcelona, Spain (pp. 46-55)

Ge, G., Sun, L., & Li, Y. F. (2022, October). A systematic approach to develop an autopilot sensor monitoring system for autonomous delivery vehicles based on the STPA method. In 2022 IEEE International Symposium on Software Reliability Engineering Workshops (ISSREW) (pp. 318-325). IEEE.

Gongadze, S. & Maassen, A. (2023, January). Paris' Vision for a '15-Minute City' Sparks a Global Movement. Available online at: https://www.wri.org/insights/paris-15-minute-city. Last accessed on: 30-06-2023.

Grieser, J., Zhang, M., Warnecke, T., & Rausch, A. (2020, August). Assuring the safety of end-to-end learning-based autonomous driving through runtime monitoring. In 2020 23rd Euromicro Conference on Digital System Design (DSD) (pp. 476-483). IEEE.

Kwon, J., Mugabarigira, B. A., & Jeong, J. P. (2022, January). Web-based car control and monitoring for safe driving of autonomous vehicles. In 2022 International Conference on Information Networking (ICOIN) (pp. 435-440). IEEE.

Li, B., Liu, S., Tang, J., Gaudiot, J. L., Zhang, L., & Kong, Q. (2020). Autonomous last-mile delivery vehicles in complex traffic environments. *Computer*, *53*(11), 26-35.

Mauritz, M. (2019). Engineering of safe autonomous vehicles through seamless integration of system development and system operation (Doctoral dissertation, Dissertation, Clausthal-Zellerfeld, Technische Universität Clausthal, 2019).

Seber, G., Czerwionka, P., Hegerhorst, T., Schappacher, M., von Bergner, A., Zhang, M., Wilken, N., Schumann, D., & Stürmer, T.. Schlussbericht VanAssist (engl.: Final report project VanAssist). Technical report, 2021.

Sherry, L., Shortle, J., Donohue, G., Berlin, B., & West, J. (2020, September). Autonomous Systems Design, Testing, and Deployment: Lessons Learned from The Deployment of an Autonomous Shuttle Bus. In 2020 Integrated Communications Navigation and Surveillance Conference (ICNS) (pp. 5D1-1). IEEE.